\documentclass[aps,prl,showpacs,twocolumn,superscriptaddress]{revtex4-2}%

\usepackage{graphicx}
\usepackage{dcolumn}
\usepackage{bm}

\usepackage{etex}

\usepackage{float}  

\usepackage{graphicx}
\usepackage{graphics}
\usepackage[caption=false]{subfig}
\usepackage{color}
\usepackage[english]{babel}
\usepackage[utf8]{inputenc}
\usepackage{ae}
\usepackage{latexsym}
\usepackage{dcolumn}
\usepackage{bm}
\usepackage[table]{xcolor}
\usepackage{array}
\usepackage{url}
\usepackage{epsf}
\usepackage{epsfig}
\usepackage{pstricks,pst-grad}
\usepackage[pdftex,colorlinks]{hyperref}
\usepackage{amsmath}
\usepackage{amssymb}
\usepackage{ragged2e}
\usepackage{comment}

\begin{document}

\title{Quantifying the Critical Micelle Concentration of Nonionic and Ionic Surfactants by Self-Consistent Field Theory}

\author{Chao Duan}
\affiliation{Department of Chemical and Biomolecular Engineering, University of California Berkeley, CA 94720, USA}

\author{Mu Wang}
\affiliation{Strategic Innovation \& Technology, The Procter \& Gamble Company, Cincinnati, OH 45217, USA}

\author{Ahmad Ghobadi}
\affiliation{Strategic Innovation \& Technology, The Procter \& Gamble Company, Cincinnati, OH 45217, USA}

\author{David M. Eike}
\affiliation{Corporate Functions Research \& Development, The Procter \& Gamble Company, Mason, OH 45040, USA}

\author{Rui Wang}
\affiliation{Department of Chemical and Biomolecular Engineering, University of California Berkeley, CA 94720, USA}
\affiliation{Materials Sciences Division, Lawrence Berkeley National Lab, Berkeley, CA 94720, USA}

\date{\today}

\begin{abstract}
Quantifying the critical micelle concentration (CMC) and understanding
its relationship with both the intrinsic molecular structures and environmental conditions are crucial for the rational design of surfactants. Here, we develop a self-consistent field theory which unifies the study of CMC, micellar structure and kinetic pathway of micellization in one framework. The long-range electrostatic interactions are accurately treated, which not only makes the theory applicable to both nonionic and ionic surfactants but also enables us to capture a variety of
salt effects. The effectiveness and  versatility of the theory is verified by applying it to three types of commonly used surfactants. For polyoxyethylene alkyl ethers
(C$_m$E$_n$) surfactants, we predict a wide span of CMC from $10^{-6}$ to $10^{-2}$M as the composition parameters $m$ and $n$ are adjusted. For the ionic
sodium dodecyl sulfate (SDS) surfactant, we show the decrease of CMC as salt concentration increases, and capture both the specific
cation effect and the specific anion effect. Furthermore, for sodium lauryl ether sulfate (SLES) surfactants, we find a
non-monotonic dependence of both the CMC and micelle
size on the number of oxyethylene groups. Our theoretical predictions of CMC are in quantitative agreement with experimental data reported in literature for all the three types of surfactants.



\end{abstract}

\maketitle

\section{1. Introduction}

Surfactants are amphiphilic molecules that contain both hydrophobic and hydrophilic groups, which play a vital role in a wide range of industrial, environmental, and biological applications due to their unique ability to alter surface/interfacial properties. \cite{Karsa2006,Rosen2012,Myers2020}
Their enrichment at the interface between two immiscible media, such as oil and water, can greatly reduce surface/interfacial tension. \cite{Chanda_2006,Wu_2016,Bergfreund_2021}
This is essential for processes like emulsification, detergency, foaming and compatibilization, making surfactants indispensable in commodity products from household cleaners, cosmetics to pharmaceuticals. \cite{Falbe1987,UCHEGBU19951,PUGH199667,Scheibel_2004,Liu_2006,Lourith_2009,VORONOV201495,Gallou_2016}
Besides, due to their interfacial activity and self-assembly feature, surfactants are also critical in many advanced functional materials used in biotechnology and nanotechnology. \cite{Richardson_2015,Kosaric_2017,Le_Guenic_2018,Nitschke_2021}

One of the most important characteristics of the surfactants is the  critical micelle concentration (CMC), above which surfactant molecules in solution have a strong tendency to aggregate into micelles. \cite{Mukerjee2018CriticalMC,Perinelli_2020}
CMC is accompanied by the occurrence of the  turning point in the surface/interfacial tension. Below CMC, surfactant molecules
primarily reside at the surface/interface, reducing the tension. As CMC is reached, the surface/interfacial tension reduces to a minimum, which remains almost constant upon further addition of surfactants. CMC is thus widely adopted as a metric to evaluate the surface/interfacial activity of surfactants. Furthermore, the formation of micelles after CMC also significantly alters solution properties, influencing the distribution and functionality of other compounds, such as dyes, perfumes and drug molecules. \cite{Lawrence_1994,Tehrani_Bagha_2013,BRADBURY2016220}
For ionic surfactants, it is found that both CMC and micelle structures are very sensitive to ionic environment, e.g., pH, salt concentration, valency and chemical identity of ions. \cite{Belhaj2020,Sarkar_2021}
Therefore, quantifying CMC and understanding its relationship with both the molecular structure and environmental conditions is crucial to ensuring the desired effectiveness and functionality of the surfactants.  

While experimental techniques such as tensiometry have been greatly developed \cite{adamson1997physical,PARIA200475,Tanhaei2013}, computational methods provides an alternative tool to predict surface/interfacial tension in a high-throughput and low-cost way. Many efforts have been made to model surfactant micellization since the pioneering work of Blankschtein, Nagarajan, Ruckenstein and Israelachvili. \cite{Puvvada_1990,Nagarajan_1991,Puvvada_1992,Israelachvili_1994,Shiloach_1998,Mulqueen_1999,Mulqueen_2000,Nagarajan_2000,Mulqueen_2001,Nagarajan_2001}
Earlier thermodynamic theories constructed the free energy of micellization based on shape parameters which are determined from data fitting. \cite{Nagarajan_1991,Nagarajan_2001}
These phenomenological models lack explicit connection to the detailed molecular structures and are difficult to be applied to ionic surfactants.
Field theoretical approach, originally derived for polymeric systems, provides an easier access to the surfactant structure. \cite{Lauw_2003,Lokar_2004,Postmus_2008b,SHEN202384,Li_2024}
Nguyen et al. developed a molecularly informed field theory to predict the CMC of biologically based protein surfactants. \cite{Nguyen_2023}
The coarse-grained parameters combining both the charge and non-charge interactions between species are informed from an separated molecular dynamic (MD) simulation on two short fragments of a protein. Molecular simulations are also widely used to study the micellization process.
Anderson et al. performed dissipative particle dynamics (DPD) simulations to predict CMC and aggregation number for a series of commonly-used surfactants. \cite{Anderson_2017,Anderson_2018,Anderson_2023}
They adopted a systematic thermodynamic scheme to obtain interaction parameters from comparison with experimentally measured partition coefficients. However, simulation approaches usually suffer high computational cost as a result of large simulation box required to sample a sufficient number of micelles. This is especially challenging for systems of very low CMC or very large critical micelles.
Furthermore, machine learning-based methods arouse great interest in recent years. \cite{CHEN2024135276,Brozos_2024}
Using graph convolutional neural networks, Qin et al. predicted CMC for different types of surfactants. \cite{Qin_2021} As mentioned by the authors, this approach is limited by the availability and accuracy of the experimental data used to train the neural networks. Molecular insights, such as micellar structure and specie distribution, are also difficult to be extracted by this approach. To our knowledge, effectively predicting CMC for a wide range of surfactants and surfactant mixtures remains a great challenge.

In this work, we develop a molecular theory which systematically incorporates the self-consistent field calculation of micellar structure and free energy into the dilute solution thermodynamics. Our theory is applicable to both nonionic and ionic surfactants, and can be easily generalized to surfactant mixtures. Long-range electrostatic interactions are accurately treated and decoupled from the short-range van der Waals interactions. This advantage allows us to capture a variety of salt effects like counterion binding, salt concentration dependence and specific ion effect. \cite{Kralchevsky_1999,Jungwirth_2005}
Molecular features of the surfactant, such as composition, architecture and charge pattern, are also explicitly considered. The studies of CMC, micellar structure and kinetic pathway of micellization are included in a unified framework. We apply the theory to quantify CMC of three types of commonly used surfactants, polyoxyethylene alkyl ethers (C$_m$E$_n$), sodium dodecyl sulfate (SDS) and sodium lauryl ether sulfate (SLES). Our theoretical predictions of CMC are in quantitative agreement with the experimental data reported in literature.

\section{2. Theory}

\begin{figure}[H]
\centering
\includegraphics[width=0.48\textwidth]{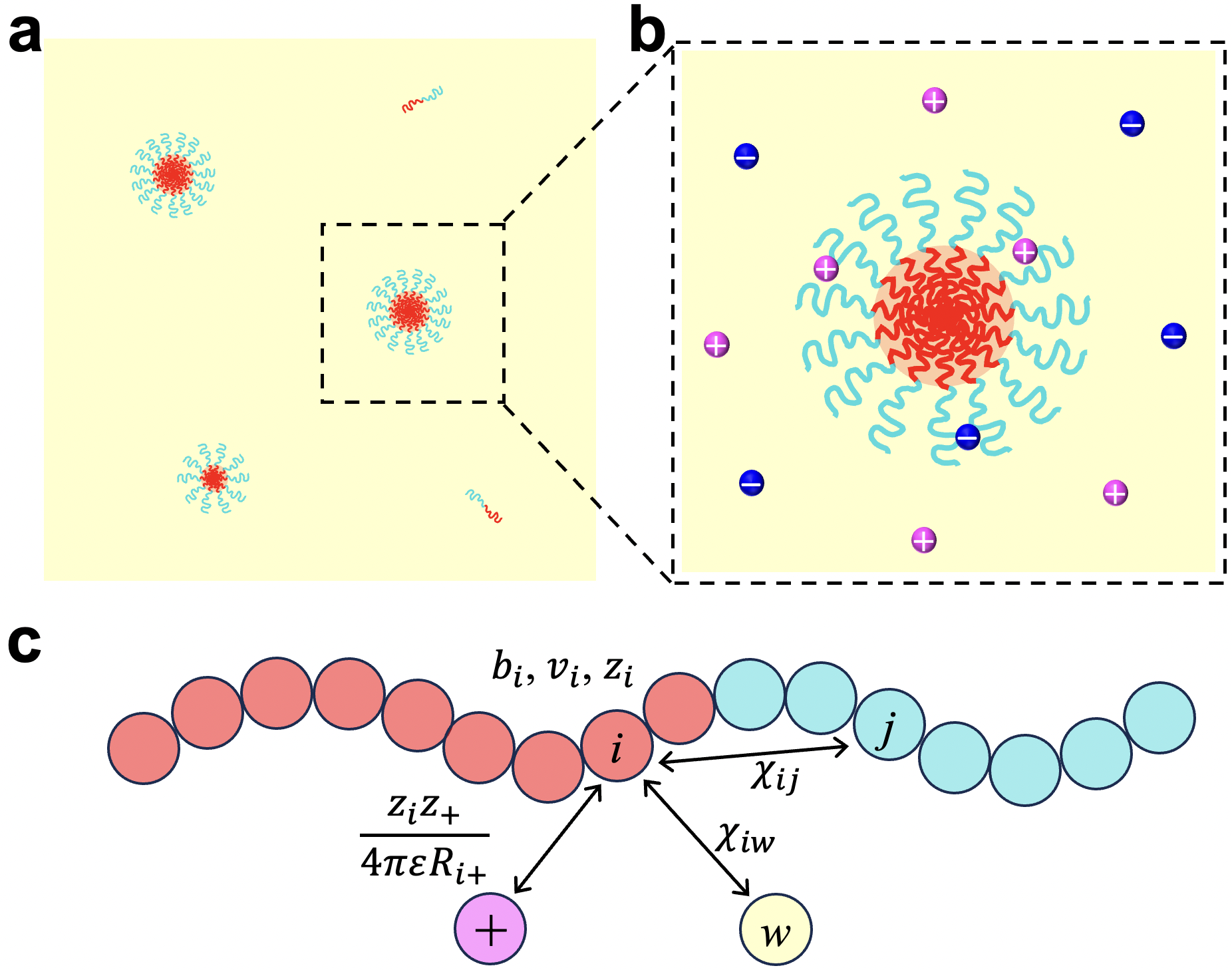}
\caption{
(a) Schematic of the total system consisting of free surfactant molecules and micelles with different aggregation number. (b) A subsystem containing one isolated micelle in the presence of salt ions. (c) The coarse-grained model for a surfactant molecule. Segment $i$ has a length $b_i$, a volume $v_i$ and a charge with valency $z_i$. The non-electrostatic interaction between the $i$-th segment and the $j$-th segment is described by the Flory-Huggins parameter $\chi_{ij}$, whereas that between the $i$-th segment and the solvent is described by $\chi_{iw}$.  
}
\label{Schematic}
\end{figure}

As illustrated in Figure \ref{Schematic}a, we consider a system consisting of spherical micelles with different aggregation number $k$ (termed $k$-micelle hereafter). $k=1$ specifies free surfactants in the solution. To effectively describe micelles, we focus on a subvolume containing only one isolated $k$-micelle (see Figure \ref{Schematic}b). The self-consistent field theory (SCFT) is applied to the subvolume to calculate the morphology and free energy of such micelle. This
information is then incorporated into the framework of dilute
solution thermodynamics to reconstruct the entire solution.

\subsection{2.1 Self-Consistent Field Theory for an Isolated Micelle}
As shown in Figure \ref{Schematic}b, we focus on an isolated micelle assembled by $k$ surfactants in a subvolume of the entire solution. The subvolume also contains $n_w$ solvent molecules in the presence of $n_{\pm}$ mobile ions. The subvolume is taken as a semicanonical ensemble \cite{Chang_2006,WangJF_2010,Wang_2012,Wang_2014,Mysona2019,Duan_2020,Duan:2023vk,Duan_2024}: the number of surfactants is fixed whereas solvent and mobile ions are connected with a bulk salt solution of ion concentration $c^b_{\pm}$ that maintains the chemical potentials of the solvent $\mu_w$ and ions $\mu_{\pm}$.
We adopt the freely jointed chain model to describe surfactant molecules, which partially captures the local rigidity and imposes the restriction on the total extensibility due to the fixed bond length.
Freely jointed chain is thus more realistic than continuous Gaussian chain in modeling surfactant molecules, particularly for those with small molecular weight. \cite{Matsen_2012,Matsen_2020}
As illustrated by Figure \ref{Schematic}c, each surfactant molecule is composed of $N$ segments. The freely jointed chain model enables us to explicitly treat the molecular features of individual segments. The mapping procedure from the exact chemical structure of the surfactant to the coarse-grained segments will be introduced later in the parameterization subsection. Segment $i$ ($i$ from 1 to $N$) has a length $b_i$, volume $v_i$ and a charge with valency $z_i$. $z_i=0$ specifies the charge neutral segments. The hydrophobic interactions between different segments as well as between segment and solvent are described by Flory-Huggins (F-H) parameters. Specifically, $\chi_{ij}$ is the F-H parameter between the $i$-th segment and the $j$-th segment; $\chi_{iw}$ is that between the $i$-th segment and solvent. Mobile ions are taken as point charges with valency $z_{\pm}$.

The semicanonical partition function can be written as
\begin{align}\label{PartitionFunc}
&\Xi = \frac{1}{k!(\prod^N_{i=1}v_i)^k} 
\prod_{\alpha=1}^{k} \int \hat{D}\{ {\bf R}_{\alpha} \}
\sum_{{n_\gamma}=0}^{\infty} \prod_{\gamma}
\frac{e^{\mu_\gamma n_\gamma}}{n_\gamma! v_\gamma^{n_\gamma}} \nonumber\\
&\cdot \prod_{j=1}^{n_\gamma} \int {\rm d}{\bf r}_{\gamma,j}  \prod_{\bf r}\delta \left[ \sum^N_{i=1}\hat{\phi}_i \left(\bf r \right)+\hat{\phi}_w \left(\bf r \right) -1 \right]{\rm exp}\left(-H \right)
\end{align}
where $\gamma=w,\pm$ represents all the small molecules with volume $v_\gamma$. For simplicity, we follow the treatment of Ginzburg et al. by taking the volume of solvent as the reference and assuming $v_i=v_w=v$. \cite{Ginzburg_2011}
$\int \hat{D}\{ {\bf R}_{\alpha} \}$ denotes integration over all chain configurations weighted by the statistics of freely jointed chain. $\hat{\phi}_i (\bf{r})$ and $\hat{\phi}_w (\bf{r})$ are the instantaneous volume fractions of segment $i$ and solvent, respectively. The $\delta$ functional accounts for the incompressibility. The Hamiltonian $H$ in Eq. \ref{PartitionFunc} is given by
\begin{align}\label{Hamiltonian}
H=&\frac{1}{v} \int {\rm d}{\bf r}
\sum^N_{i=1} \hat{\phi}_i({\bf r}) \left[ \sum^N_{j \ge i} \chi_{ij}  \hat{\phi}_j({\bf r})
+\chi_{iw} \hat{\phi}_w({\bf r}) \right]  \nonumber\\
&+\frac{1}{2} \int {\rm d}{\bf r} {\rm d}{\bf r}^{\prime}
\hat{c}_e({\bf r}) C({\bf r},{\bf r}^{\prime}) \hat{c}_e({\bf r}^{\prime})
\end{align}
It includes two contributions: 1) short-range hydrophobic interactions between different segments as well as between segments and solvent; and 2) long-range electrostatic interactions between all charged species. $\hat{c}_e({\bf r})=\sum_{k=\pm} z_k \hat{c}_k({\bf r})+\sum^N_{i=1}z_i\hat{\phi}_i({\bf r})/v$ is the net charge density. $C({\bf r},{\bf r}^{\prime})$ is the Coulomb operator, satisfying $-\nabla \cdot [\varepsilon({\bf r})\nabla C({\bf r},{\bf r}^{\prime}) ] =\delta({\bf r}-{\bf r}^{\prime})$. $\epsilon({\bf r})=kT \epsilon_0 \epsilon_r({\bf r}) /e^2$ is the scaled permittivity with $\epsilon_0$ the vacuum permittivity, $e$ the elementary charge and $\epsilon_r({\bf r})$ the local dielectric constant. $\epsilon_r({\bf r})$ can be evaluated based on the local composition through a specific mixing rule (e.g. Clausius-Mossotti relation). \cite{Wang_2011,Nakamura:2012aa,Zhuang_2021,Duan_2024}
It can be clearly seen that the long-range charge interactions are explicitly treated in our theory. This is a major advantage compared to previous work where the charge interactions are absorbed to an effective short-range Flory-Huggins interactions. \cite{SHEN202384,Nguyen_2023}

Following the standard field theoretical approach \cite{Fredrickson_Book2006,Fredrickson_Book2023}, the key results of the SCFT are the following set of equations for segment density $\phi_i({\bf r})$, solvent density $\phi_w({\bf r})$, their conjugate
fields $\omega_i(\bf r)$ and $\omega_w(\bf r)$, electrostatic potential $\psi({\bf r})$ and ion concentration $c_{\pm}({\bf r})$:
\begin{subequations}\label{SCF_Eqs}
\begin{align}\label{wi}
\omega_i({\bf r})& = \chi_{iw}\phi_w({\bf r})+\sum^N_{j=1}\chi_{ij}\phi_j({\bf r}) +\xi({\bf r}) +z_i\psi({\bf r})
\end{align}
\begin{align}
\omega_w({\bf r})& = \sum^N_{i=1}\chi_{iw}\phi_i({\bf r}) +\xi({\bf r})
\end{align}
\begin{align}
\phi_i({\bf r}) &= \frac{k}{Q_s} q({\bf r};i)
e^{\omega_i({\bf r})}
q^{\dagger}({\bf r};i)
\end{align}
\begin{align}
\phi_w({\bf r}) &= e^{\mu_w-\omega_w({\bf r})}
\end{align}
\begin{align}
-\nabla\cdot[\epsilon({\bf r})\nabla\psi({\bf r})] &=\sum_{k=\pm} z_k c_k({\bf r})+\sum^N_{i=1}\frac{z_i\phi_i({\bf r})}{v}
\end{align}
\begin{align}\label{c_ion}
c_{\pm}({\bf r}) &=\lambda_{\pm} e^{- z_{\pm} \psi({\bf r})}
\end{align}
\end{subequations}
where $\xi({\bf r})$ is the incompressibility pressure. $\lambda_{\pm}=e^{\mu_{\pm}}/v_{\pm}$ is the fugacity of the ions controlled by the bulk salt concentration. Eqs. \ref{wi}-\ref{c_ion} are derived in the mean-field framework which cannot capture the effects of spatial varying dielectric medium and the ion-ion correlation as a consequence of the fluctuation of the electrostatic field. To account for the local fluctuation effect, the Born solvation energy $u_{\pm}({\bf r})$ can be included into the Boltzmann factor in Eq. \ref{c_ion} as:
\begin{align}
c_{\pm}({\bf r}) =& \lambda_{\pm}e^{- z_{\pm} \psi({\bf r})-u_{\pm}({\bf r})}
\end{align}
where
\begin{align}\label{u_Born}
u_{\pm}({\bf r})=\frac{z_{\pm}^2 e^2}{8\pi \epsilon({\bf r}) a_{\pm} }
\end{align}
with $a_{\pm}$ the Born radius of cations and anions, respectively.
The inclusion of the Born solvation energy can be rigourously achieved by taking the Gaussian fluctuation of the electrostatic field and retaining the nonuniversal contribution in the length scale of the ion size. \cite{Wang:2010wk,Agrawal:2022ux,Agrawal_2024}
The Born solvation energy accounts for the electrostatic interaction between the ion and the local dielectric medium, particularly important in systems with spatially varying dielectric permittivity.

$Q_s$ is the partition function of a single surfactant molecule given by $Q_s=v^{-1}\int {\rm d}{\bf r} q({\bf r};i)e^{\omega_i({\bf r})}q^{\dagger}({\bf r};i)$. $q({\bf r};i)$ and $q^{\dagger}({\bf r};i)$ are the forward and backward chain propagators, respectively, determined by the Chapman–Kolmogorov equation \cite{Fredrickson_Book2006}:
\begin{subequations}\label{q0}
\begin{align}\label{q0_forward}
q({\bf r};i) &= e^{-\omega_i({\bf r})} \int d{\bf r^{\prime}} \Phi_i({\bf r}-{\bf r^{\prime}}) q({\bf r}^{\prime};i-1)
\end{align}
\begin{align}\label{q0_backward}
q^{\dagger}({\bf r};i) &= e^{-\omega_i({\bf r})} \int d{\bf r^{\prime}} \Phi_i({\bf r}-{\bf r^{\prime}}) q^{\dagger}({\bf r}^{\prime};i+1)
\end{align}
\end{subequations}
with the initial condition $q({\bf r};1)=e^{-\omega_1({\bf r})}$ and $q^{\dagger}({\bf r};N)=e^{-\omega_N({\bf r})}$. $\Phi_i (\bf R)$ is the transition probability between adjacent segments in the freely-jointed chain given by $\Phi_i ({\bf R})=(4\pi b^2_i)^{-1}\delta(|{\bf R}| -b_i)$.
The resulting free energy in the subvolume is then
\begin{align}\label{Fm}
F_k &= -k{\rm ln}Q_s+{\rm ln}(k!)-{\rm e}^{\mu_w}Q_w \nonumber\\
&+\frac{1}{v}\int {\rm d}{\bf r} \Biggl[ \sum^N_{i=1}  \phi_i \left(\sum^N_{j\ge i} \chi_{ij}\phi_j+\chi_{iw}\phi_w \right) \nonumber\\
&-\sum^N_{i=1} \omega_i\phi_i-\omega_w \phi_w \Biggl] \nonumber\\
&+\int {\rm d}{\bf r} \left [ \sum^N_{i=1}\frac{z_i\phi_i}{v}\psi-\frac{\epsilon}{2}(\nabla\psi)^2-c_+ -c_- +c^b_+ +c^b_-\right ]
\end{align}
where $Q_w=v^{-1}\int {\rm d}{\bf r} e^{-\omega_w({\bf r})} $ is the partition function of solvent.


\subsection{2.2 Dilute Solution Thermodynamics} The SCFT calculation yields the density profile and free energy of the $k$-micelle. Next, we incorporate this information into the framework of dilute solution thermodynamics to reconstruct the entire surfactant solution. \cite{Wang_2012,Wang_2014,Duan:2023vk,Duan_2024}
The free energy density of the solution with volume $V$, including the translational entropy of the micelles, can be written as:
\begin{align}
F/V &= \sum_k \{C_k F_k + C_k [{\rm ln} (C_k v_k)-1] \}
\end{align}
where $C_k$ is the concentration of the $k$-micelle, and $F_k$ is its free energy obtained by Eq. \ref{Fm} from SCFT calculation. $v_k$ is a reference volume which for simplicity can be taken as the volume of the $k$-micelle. $C_k v_k$ thus becomes its corresponding volume fraction $\phi_k$.   
In Eq. 6, we ignore the interactions between different micelles under the assumption of sufficiently dilute solution. By minimizing the free energy density subject to fixed total surfactant concentration $\sum^{\infty}_{k=1}{kC_k}$, we obtain the equilibrium concentration of the $k$-micelle as:   
\begin{align}\label{phi_m}
\phi_k &= \phi^k_1 \exp(-\Delta F_k)
\end{align}
$\phi_1$ is the fraction of the free surfactant. $\Delta F_k=F_k-kF_1$ is the formation energy of the $k$-micelle from $k$ free surfactant molecules. 

It should be noted that the determination of $F_1$, the energy of the free surfactant, depends on whether it can form a single-chain micelle. Usually, for surfactants with large molecular weight of the hydrophobic block, micelle can be formed even by a single surfactant. The calculation of $F_1$ can thus be performed using the same procedure as we calculate micelles with larger aggregation numbers. On the other hand, if a free surfactant is too short to form a single-chain micelle, $F_1$ can be evaluated by applying Eq. \ref{Fm} in a homogeneous and disordered phase under the assumption of $\phi_1 \ll 1$, which yields:
\begin{align}\label{F1}
F_1 &= 1 -N -\ln{N} + \sum^N_{i=1}\chi_{iw} 
\end{align}

Futhermore, we can absorb the factor $\phi_1^k$ in Eq. \ref{phi_m} into the work for micelle formation that includes the translational entropy by defining the grand free energy $G_k$ as:
\begin{align}\label{Gm}
G_k &= F_k-k\mu =\Delta F_k-k{\rm ln}\phi_1
\end{align}
where $\mu=F_1+{\rm ln}\phi_1$ can be interpreted as the chemical potential of the free surfactant which also remains the same within all micelles as a result of chemical equilibrium. $G_k$ can be used to track the kinetic pathway of the micellization process if we take the aggregation number as the reaction coordinate. \cite{Duan:2023vk}
Compared to the previous work by Nguyen et al. performed purely in the grand canonical ensemble, \cite{Nguyen_2023} our method can provide additional information on the equilibrium distribution of different micelles and their aggregation kinetics.

\subsection{2.3 Determination of the Critical Micelle Concentration} 
The equilibrium distribution of micelles with different aggregation numbers obtained in the dilute solution thermodynamics enables us to determine the critical micelle concentration (CMC). CMC is the concentration above which surfactant molecules have a strong tendency to aggregate into micelles. Here, we adopt the definition of CMC such that the total volume fraction of all micelles in the solution equals that of the free surfactants, i.e. $\phi_{1,CMC}=\sum^{\infty}_{k = k^*} \phi_{k,CMC}$. $k^*$ is the minimum aggregation number that a multi-surfactant micelle can be formed. Therefore, the total volume fraction of surfactants at CMC is:
\begin{align}\label{CMC}
\phi_{\rm CMC}=\phi_{1,{\rm CMC}}+\sum^{\infty}_{k =k^*} \phi_{k,{\rm CMC}} =2\phi_{1,{\rm CMC}}
\end{align}
which can be straightforwardly converted to the CMC in terms of the molar concentration as $c_{\rm CMC}=\rho_s \phi_{{\rm CMC}} /M_s$, with $\rho_s$ and $M_s$ the mass density and molecular weight of the surfactant, respectively.

\begin{figure}[H]
\centering
\includegraphics[width=0.45\textwidth]{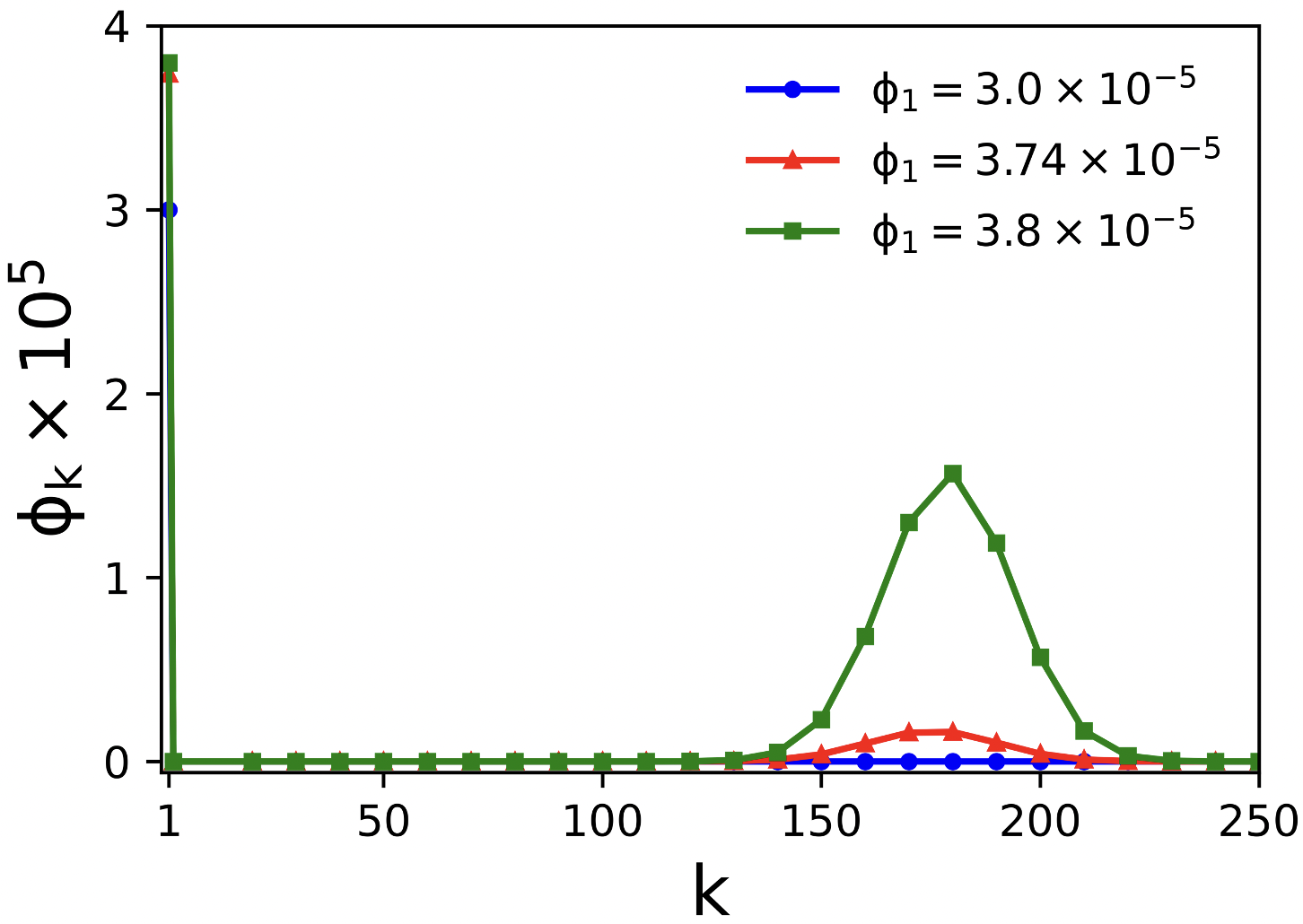}
\caption{The illustration of determining the critical micelle concentration (CMC). The equilibrium distributions of micelles is plotted as a function of the aggregation number $k$ for octaethylene glycol monododecyl ether (C$_{12}$E$_8$) surfactant with different free surfactant concentration $\phi_1$. $\phi_{1,{\rm CMC}}=3.74\times 10^{-5}$ is identified according to $\phi_{1,{\rm CMC}} = \sum^{\infty}_{k =k^*} \phi_{k,{\rm CMC}}$. CMC is thus $\phi_{\rm CMC}=2\phi_{1,{\rm CMC}}=7.48 \times 10^{-5}$.
}
\label{CMC_illust}
\end{figure}

Figure \ref{CMC_illust} illustrate the determination of the CMC for octaethylene glycol monododecyl ether (C$_{12}$E$_8$) surfactant, by plotting the equilibrium distribution of micelles with increasing free surfactant concentration. When $\phi_1$ is small ($3.0\times10^{-5}$), micelles can hardly be formed as a result of large translational entropy of free surfactant. When $\phi_1$ reaches a critical concentration ($3.74\times 10^{-5}$), a clear distribution of large micelles can be observed, and the sum of their concentration exactly equals that of the free surfactant. This allows us to identifies the CMC. The distribution has a maximum at $k=180$, which can be identified as the critical aggregation number. Beyond CMC, a slight increase of $\phi_1$ ($3.8\times 10^{-5}$) leads to a dramatic increase of the micelle concentration, where the system is dominated by large micelles.       

We note that other definitions of CMC are also used in literature. One can define it by equalizing the grand free energies between the free surfacant and the micelle with the critical aggregation number. However, CMC determined by different definitions are qualitatively the same, just with slight quantitative difference.

\subsection{2.4 Parameterization}
We follow the approach developed by Ginzburg et al. which uses water as the reference of a single segment. \cite{Ginzburg_2011}
The number of segments in any other molecule or block is determined by dividing its molecular weight by 18, the mass of the water molecule. Therefore, $v_i = v_w=18{\rm mL}/{\rm mol}=0.03{\rm nm}^3$. The Kuhn length $b_i$ is set to be 0.9nm for flexible blocks, which comes from the best fit to the interfacial tension of the dodecane/water mixture in the literature.

To determine the Flory-Huggins parameters between different segments $\chi_{ij}$ as well as between segments and solvent $\chi_{iw}$, we start from the simplist system of polyoxyethylene alkyl ether (C$_m$E$_n$) surfactants. Ginzburg et al. calibrated the $\chi$ parameters by fitting the experimental data of interfacial tension. \cite{Ginzburg_2011}
We will first confirm that the values of such parameters are also valid in predicting the CMC. Then we expand our calibration to surfactants with chemical composition close to the reference C$_m$E$_n$, which differs from C$_m$E$_n$ only by one building block. Because the $\chi$ parameters shared by the C$_m$E$_n$ system have already been known, we only need to calibrate a few $\chi$ parameters relevant to the additional block via fitting to the reported CMC data from experiments. This largely reduces the difficulty and uncertainty of the fitting. Next, we continue performing this procedure to other surfactants built upon the information of the calibrated surfactants. The success of this process will gradually expands the library of the reference surfactants by adding newly calibrated $\chi_{ij}$ and $\chi_{iw}$.

Specifically, we start from the C$_m$E$_n$ surfactants in water (W), where $\chi_{\rm CE}$, $\chi_{\rm CW}$, and $\chi_{\rm EW}$ are adopted from the work of Ginzburg et al. \cite{Ginzburg_2011}
Next, we turn to sodium dodecyl sulfate (SDS) surfactant, which consistes of hydrophobic alkyl block (C) and a charged hydrophilic sulfate group (S). This leads to two unknown parameters, $\chi_{\rm CS}$ and $\chi_{\rm SW}$, which will be calibrated by fitting the experimental CMC data of SDS. Furthermore, combining the information of C$_m$E$_n$ and SDS, we can proceed to a more complex surfactant, Sodium lauryl ether sulfate  (SLES). SLES is a derivative of SDS with an additional ethoxy (E) block. Only a single unknown parameter $\chi_{\rm ES}$ needs to be calibrated for SLES. Following this parameterization procedure, our theory can be generalized to a variety of nonionic and ionic surfactants.

To accelerate the data fitting process, we can use the solubility parameters reported in the literature as an initial guess, i.e. $\chi_{ij} = v (\delta_i - \delta_j)^2 /kT $ where $\delta_i$ is the Hildebrand solubility parameter of segment $i$. \cite{Barton_2017}
Another approach to roughly estimate the parameter is by mapping $\chi_{ij}$ to the interaction parameters obtained by molecular simulations. \cite{Groot_1997,Ginzburg_2011,Anderson_2017}
We also want to note that our theory is able to describe self-assembled aggregates with different shapes, such as spherical and cylindrical micelles and membranes (manifested as large vesicles). Here we focus on the spherical micelles for the numerical convenience.

\section{3. Results and Discussion}

Our theory is applicable to a variety of nonionic and ionic surfactants. Here, we validate the effectiveness of our theory by applying it to three commonly used surfactants: C$_m$E$_n$, SDS and SLES.

\subsection{3.1 Polyoxyethylene Alkyl Ether (C$_m$E$_n$) Surfactants}
\begin{figure}[H]
\centering
\includegraphics[width=0.48\textwidth]{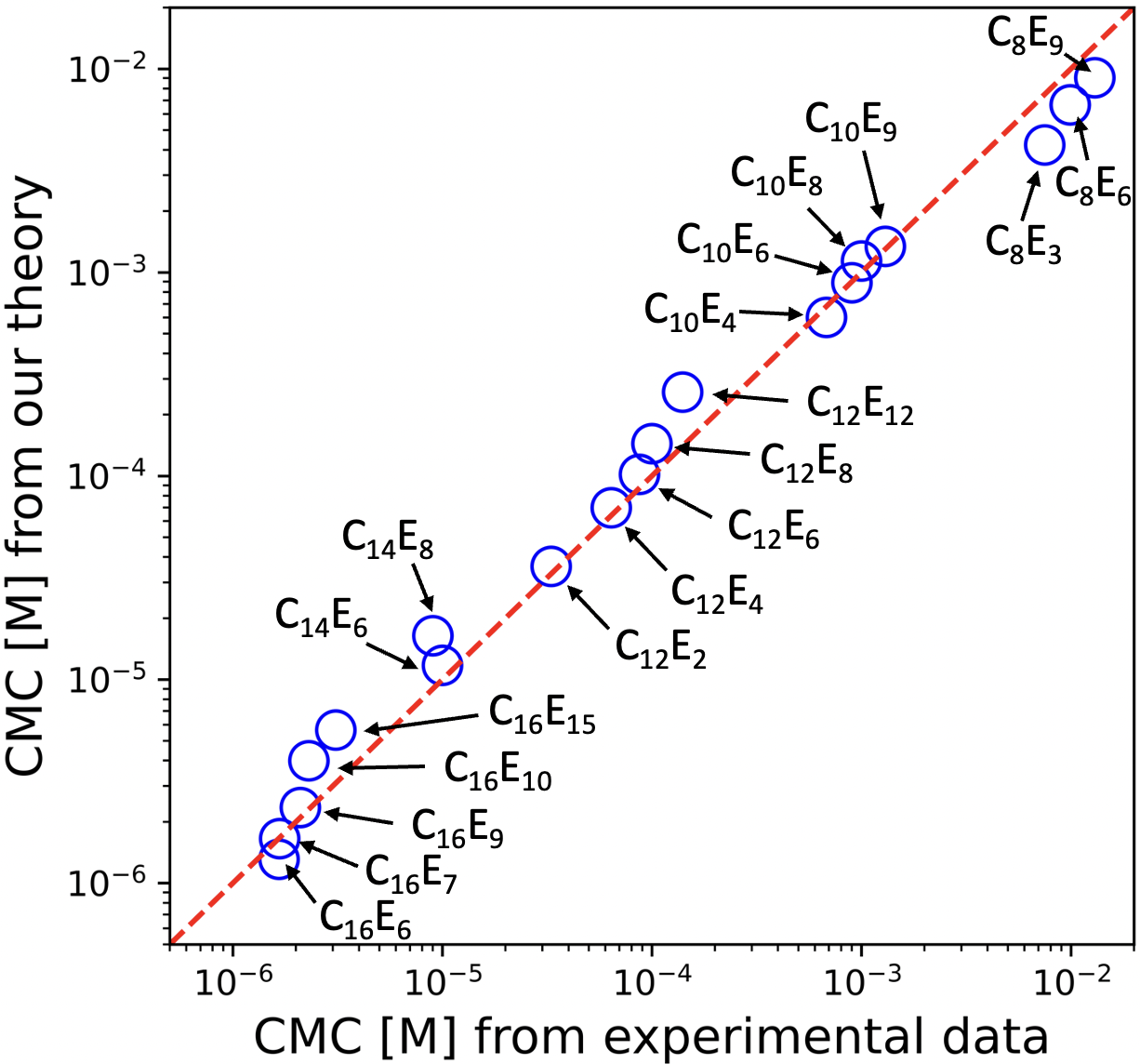}
\caption{
Comparison of the CMC predicted by our theory ($y$ axis) with the experimental data ($x$ axis) reported in Ref. \cite{Huibers_1996} for a series of C$_m$E$_n$ surfactants. The dashed diagonal indicates the exact agreement between theory and experiments. $\chi_{\rm CE}=0.25$, $\chi_{\rm CW}=2.47$, and $\chi_{\rm EW}=0.09$, are taken from Ref. \cite{Ginzburg_2011}
}
\label{CMC_CmEn}
\end{figure}

Polyoxyethylene alkyl ethers are made up of an alkyl chain (C block) with $m$ methylene groups and a hydrophilic part (E block) with $n$ oxyethylene units. \cite{BERTHOD200169}
They are non-ionic surfactants with excellent emulsifying, solubilizing, and wetting properties. \cite{YADA201951}
We first calculate the CMC of C$_m$E$_n$ for a wide span of composition parameters, $m=8 \sim 16$ and $n=2 \sim 15$. The Flory-Huggins interaction parameters $\chi_{\rm CE}=0.25$, $\chi_{\rm CW}=2.47$, and $\chi_{\rm EW}=0.09$ are adopted from the previous work of Ginzburg et al., which are calibrated by fitting the experimental data of interfacial tension. \cite{Ginzburg_2011}
Figure \ref{CMC_CmEn} shows the comparison between our theoretical prediction (the $y$ axis) and the experimental data reported in literature (the $x$ axis). It can be clearly seen that most of the points are very close to the diagonal. Therefore, with the model parameters calibrated in a separate system, our theory is in a quantitative agreement with experiments for a broad range of CMC from $10^{-6}$ to $10^{-2}$M. Our theory also elucidate the trend that CMC decreases as the number of methylene groups ($m$) increases, whereas it increases with more oxyethylene units ($n$). CMC is more sensitive to the change of $m$ than $n$.

\begin{figure}[H]
\centering
\includegraphics[width=0.4\textwidth]{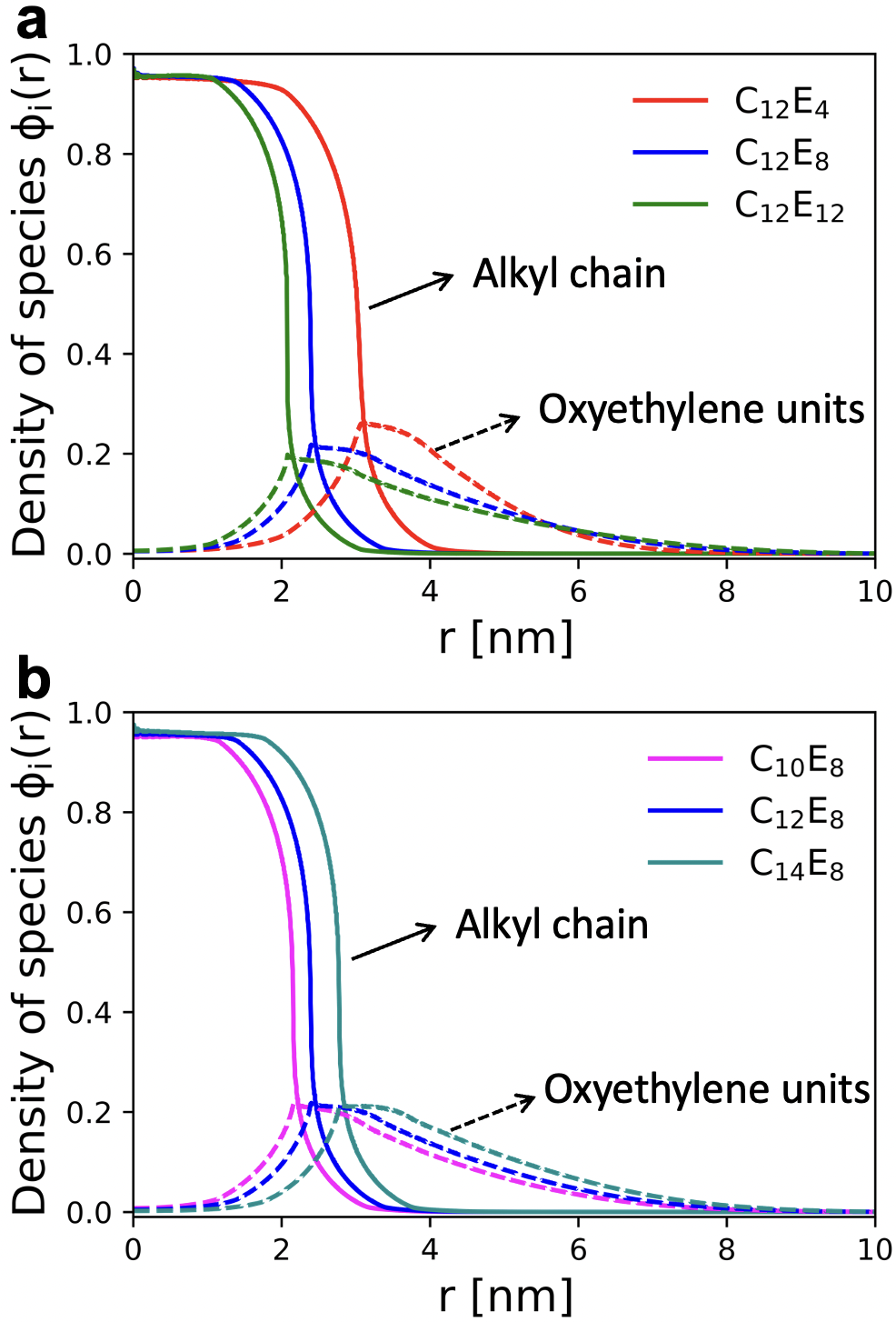}
\caption{
Microstructures of the micelles formed by C$_m$E$_n$ surfactants. Density distributions of different species $\phi_i(r)$ are plotted in the radial direction for (a) various $n$ with fixed $m=12$ and (b) various $m$ with fixed $n=8$. Solid and dashed lines represent the distributions of alkyl chain (C block) and oxyethylene units (E block), respectively.
}
\label{Profile_CmEn}
\end{figure}

Besides the quantification of CMC, our theory is also able to provide the microstructure of micelles. Figure \ref{Profile_CmEn} plots density distributions of hydrophobic methylene groups and hydrophilic oxyethylene units for various $m$ and $n$. It should be noted that there is a wide distribution of micelles with different aggregation numbers as shown in Figure \ref{CMC_illust}. Here we plot the most representative micelle structure which is formed at CMC with the critical aggregation number (i.e. the maximum in Figure \ref{CMC_illust}). As shown in Figure \ref{Profile_CmEn}, a clear ``core-shell" structure can be observed: the hydrophobic C groups form a condensed inner core while the hydrophilic E groups form a diffusive outer shell. The micelle core becomes smaller as $n$ increases (Figure \ref{Profile_CmEn}a), indicating a smaller critical aggregation number. This can be explained by the change of the packing parameter defined by $P=v_T/(a_H l_T)$, where $v_T$ and $l_T$ is the volume and the maximum extended length of the hydrophobic tail, respectively. $a_H$ is the area of the hydrophilic head at the core-shell interface. Increasing the number of hydrophilic units ($n$) leads to larger $a_H$ due to the excluded volume effect, and consequently a smaller $P$. This results in a smaller spherical micelle. Similar trend can also be observed in Figure \ref{Profile_CmEn}b as we reduce the number of hydrophobic C groups ($m$).

\subsection{3.2 Sodium Dodecyl Sulfate (SDS) Surfactants}

\begin{figure}[H]
\centering
\includegraphics[width=0.43\textwidth]{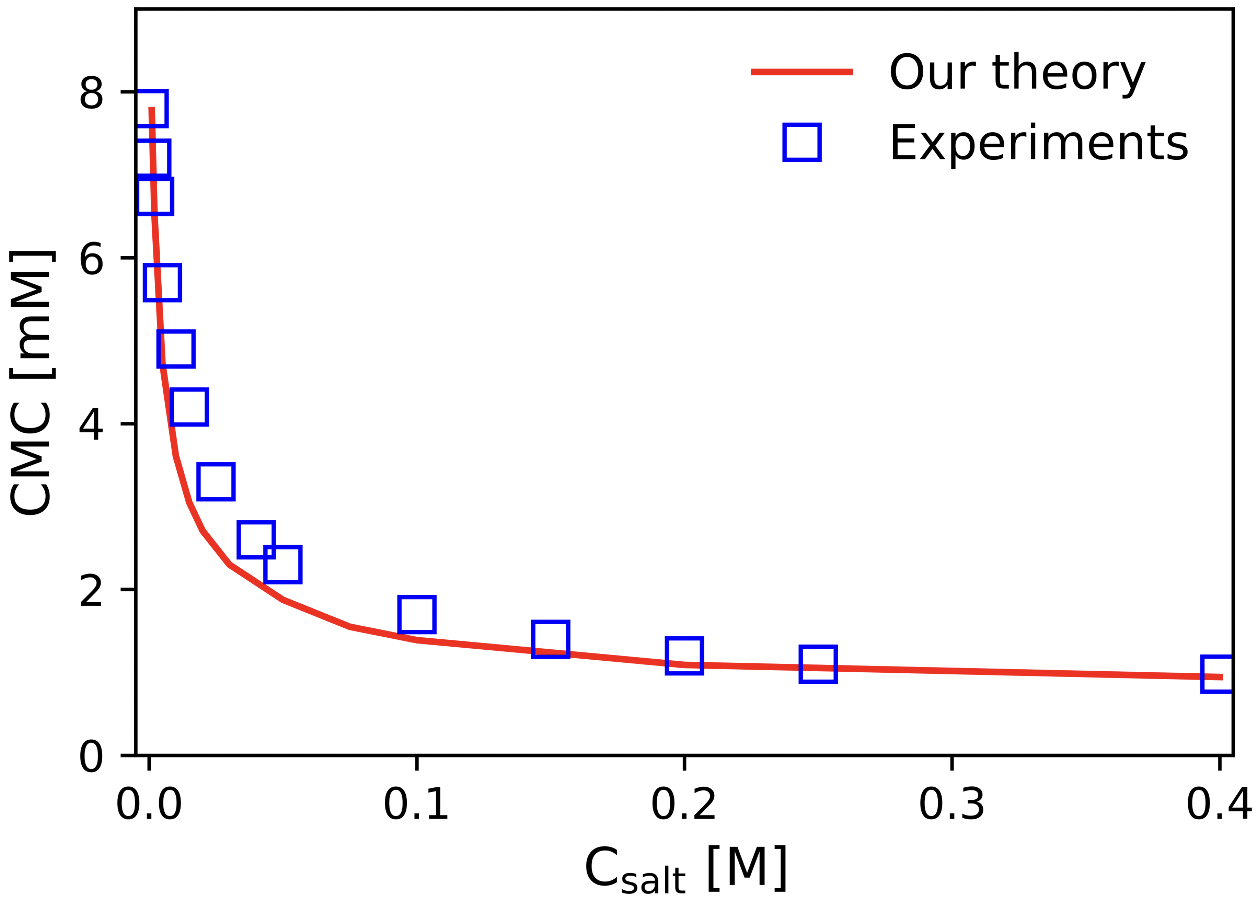}
\caption{
Comparison of the CMC predicted by our theory and experimental data reported in Ref. \cite{Dutkiewicz2002} for SDS surfactant in NaCl solutions with different salt concentrations $C_{\rm salt}$.   
The Born radii of Na$^+$ and Cl$^-$ are 1.67 and 2.02 $\mathring{\rm A}$, respectively \cite{SATHEESANBABU1999225}. The dielectric constants of water and SDS surfactant are 80.0 and 3.0, respectively. The Flory-Huggins parameter $\chi_{\rm CW}=2.47$ is taken from the C$_m$E$_n$ system, whereas $\chi_{\rm CS}=3.5$ and 
$\chi_{\rm SW}=-2.5$ are obtained from the best fit to the experimental data in the high salt concentration regime ($C_{\rm salt}>0.1$M). 
}
\label{CMC_SDS}
\end{figure}

Sodium dodecyl sulfate (SDS) is an anionic surfactant used in many cleaning and hygiene products.
Compared to nonionic surfactants, ionic surfactants have a better performance in removing ionic contaminants due to their charged hydrophilic head groups which can lift and suspend soils in micelles. \cite{Harendra_2012}
Adding salt ions in the solution provide an effective tool to tune the CMC of SDS and hence the interfacial efficacy. \cite{Sammalkorpi_2009,Naskar2013}
Figure \ref{CMC_SDS} plots the CMC of SDS in NaCl solutions as a function of salt concentration $C_{\rm salt}$, which shows a monotonic decay as $C_{\rm salt}$ increases. At high salt concentrations ($C_{\rm salt}>0.1$M) where the charge interactions are largely screened, the decrease of CMC becomes slower and eventually approaches the limit of the neutral surfactant.

In our calculation, we maintain the value of F-H interaction parameter $\chi_{\rm CW}$ between methylene groups in dodecyl chain and water the same as we used in C$_m$E$_n$ surfactants. The two unknown parameters in SDS, $\chi_{\rm CS}$ between methylene groups and hydrophilic sulfate group as well as $\chi_{\rm SW}$ between sulfate group and water, are calibrated by fitting the experimental data at high salt concentrations. In this regime, the charge interactions are weak, leading to minimum interference in fitting non-electrostatic interaction parameters. The calibrated $\chi_{\rm CS}=3.5$ and 
$\chi_{\rm SW}=-2.5$ are then applied to the calculation of CMC at lower salt concentrations to examine the validity of the theory. The negative $\chi_{\rm SW}$ is attributed to the strong affinity between sulfate group and water as a result of hydrogen bond. Figure \ref{CMC_SDS} clearly shows that, using a single a set of F-H parameters, our theory is in quantitative agreement with the experimental data in the entire regime of salt concentration. The long-range electrostatic interactions are systematically treated in our theory and decoupled from the short-range F-H interactions, which enables us to fully capture the salt concentration dependence of CMC in ionic surfactants.

\begin{figure}[H]
\centering
\includegraphics[width=0.4\textwidth]{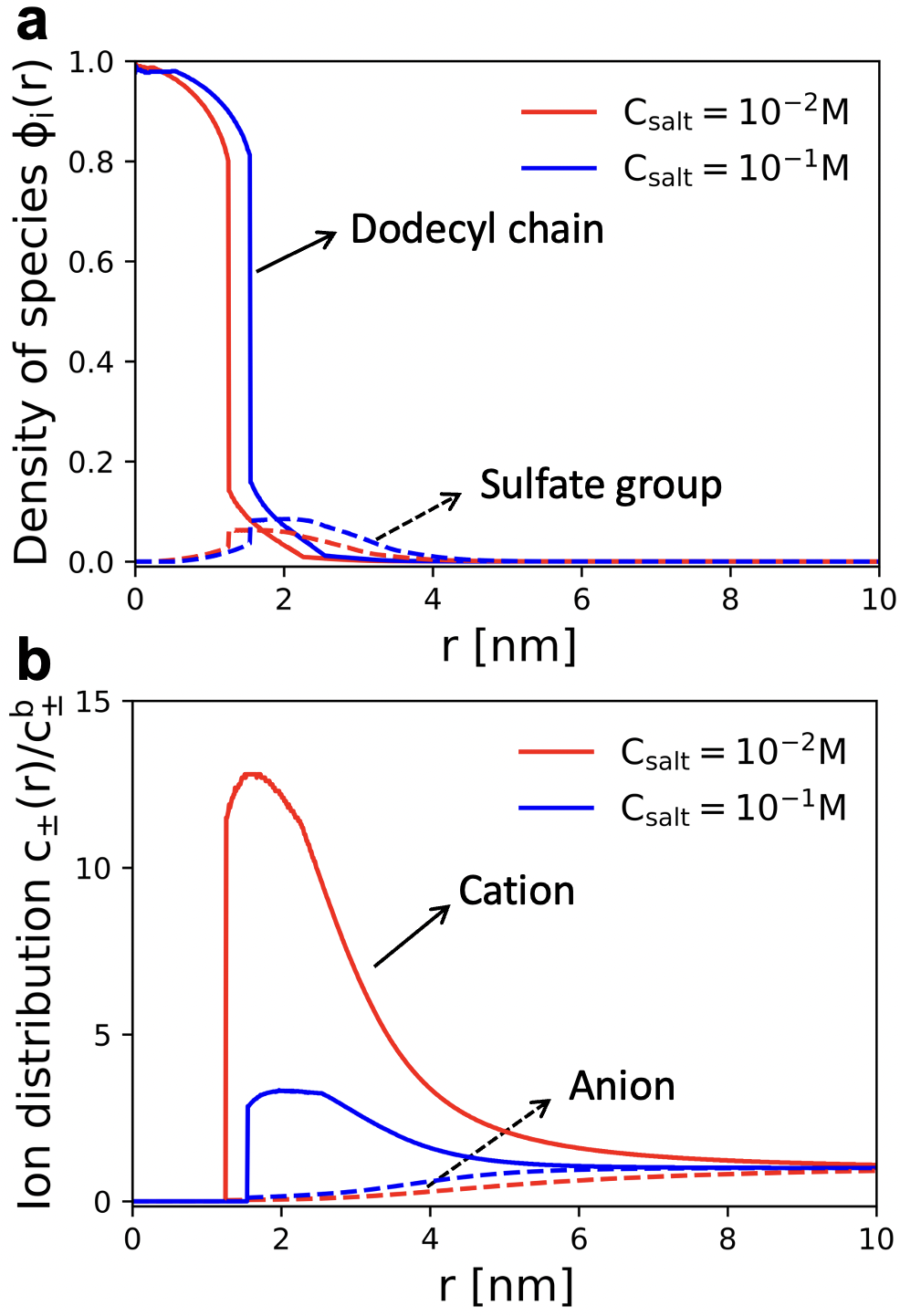}
\caption{
Microstructures of micelles formed by SDS surfactants in NaCl solutions at two different salt concentrations $C_{\rm salt}$. (a) Density profiles of dodecyl chain and sulfate group, and (b) concentrations of cation $c_+$ (solid line) and anion $c_-$ (dashed line), normalized by their corresponding bulk concentration $c^b_{\pm}$, in the radial direction.
}
\label{Profile_SDS}
\end{figure}

Figure \ref{Profile_SDS} shows the microstructure of the representative micelles (at CMC with the critical aggregation number) for SDS surfactant. The density distributions of dodecyl chain and sulfate group shown in Figure \ref{Profile_SDS}a indicates that the micelle becomes larger at higher salt concentrations. The increase of $C_{\rm salt}$ reduces the repulsive force between negatively charged sulfate groups due to the stronger screening effect. This leads to a decrease of the effective area of the hydrophilic head at the core-shell interface $a_H$ and hence increases the packing parameter $P$. Figure \ref{Profile_SDS}b plots the distribution of counterions (Na$^+$) and coions (Cl$^-$) in the vicinity of the micelle. The hydrophobic micelle core is a condensate of hydrocarbon chains which usually have very low dielectric constant $\epsilon_r <5$. Ions are excluded from this region as a result of the dramatic increase of the Born solvation energy as indicated by Eq. \ref{u_Born}. Counterions are accumulated in the hydrophilic shell around the sulfate group. The counterion concentration in the micelle shell in excess of the bulk value can be used to evaluate the counterion binding fraction. As shown in Figure \ref{Profile_SDS}b, the counterion binding fraction is higher at low salt concentrations because of the stronger electrostatic attraction from the sulfate group.

\begin{figure}[H]
\centering
\includegraphics[width=0.4\textwidth]{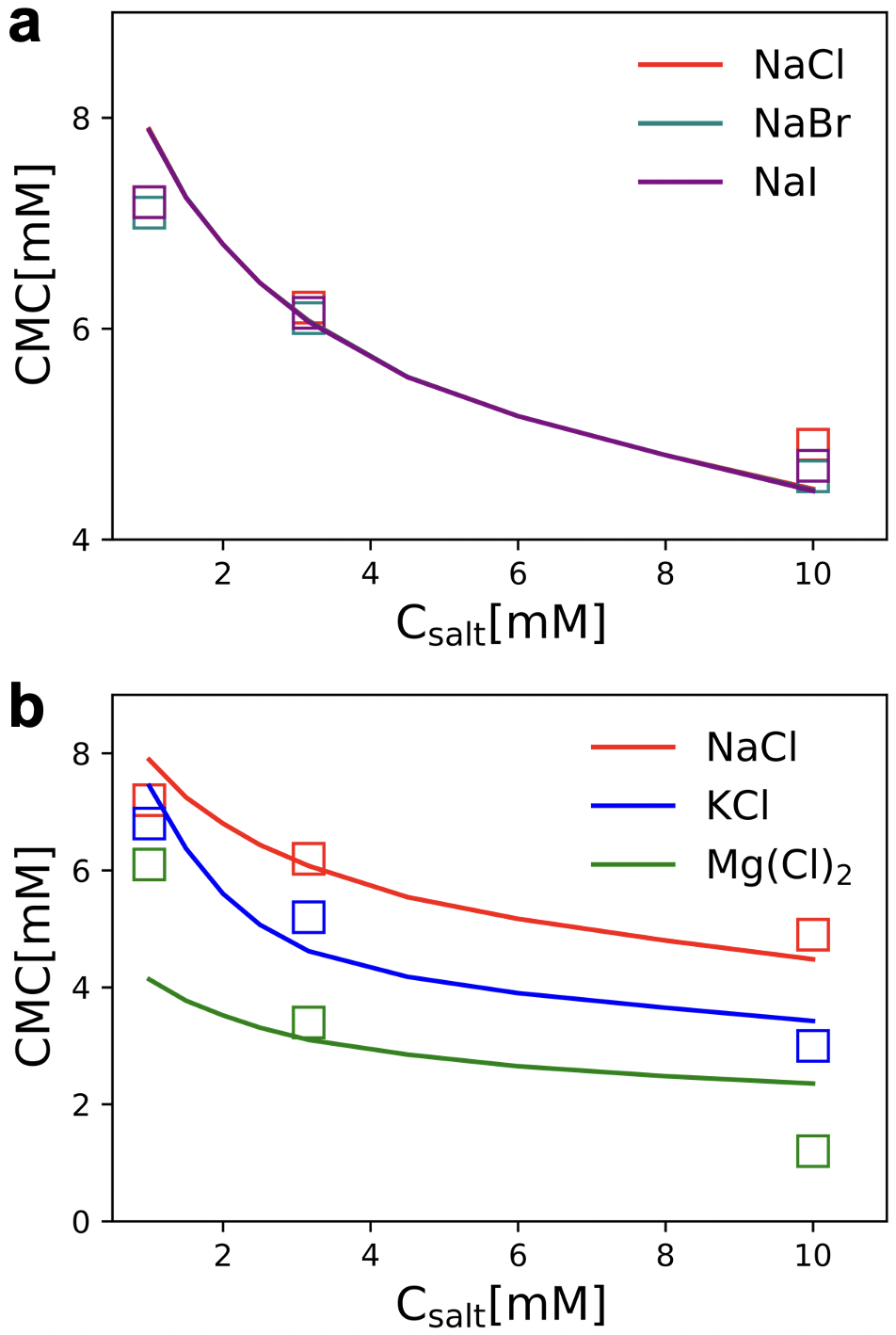}
\caption{
(a) Specific anion effect and (b) specific cation effect on the CMC of SDS surfactants. Our theoretical predictions (lines) are compared with the experimental data (symbols) reported in Ref. \cite{Martinez_de_la_Ossa_1987}. The Born radii of Na$^+$, K$^+$, Mg$^{2+}$, Cl$^-$, Br$^-$, and I$^-$ are 1.67, 2.03, 1.37, 2.02, 2.16, and 2.45 $\mathring{\rm A}$, respectively, taken from Ref. \cite{SATHEESANBABU1999225}. The dielectric constant of water and SDS surfactant are 80.0 and 3.0, respectively. All the Flory-Huggins parameters are the same as those calibrated in the NaCl solution in Fig. \ref{CMC_SDS}.
}
\label{SpecificIon}
\end{figure}

The explicit treatment of the electrostatic interactions in our theory and the inclusion of the Born solvation energy of ions also allows us to capture the dependence of CMC on the chemical identity of ions, usually known as the specific ion effect. Figure \ref{SpecificIon} plot  the CMC of SDS in different salt solutions. Our theoretical predictions are in good agreement with the experimental data for both the anion effect (Figure \ref{SpecificIon}a) and cation effect (Figure \ref{SpecificIon}b). It should be noted that the Born radii of different ions are not fitting parameters but adopted from literature. As shown in Figure \ref{SpecificIon}a, the CMC of SDS is not sensitive to sodium salt solutions with different halogen anions. The results for Cl$^-$, Br$^-$ and I$^-$ are almost indistinguishable, confirmed by both theory and experiments. As the role of coion with respect to the negatively-charged SDS, anions have negligible concentration in the vicinity of micelles, indicated by the ion distribtuion in Figure \ref{Profile_SDS}b. Anions thus have little influence on either the micelle structure or the CMC of SDS surfactant.

In stark contrast, the CMC of SDS exhibits a strong cation effect as shown in Figure \ref{SpecificIon}b. With the same Cl$^-$ anion, the CMC significantly decreases as Na$^+$ is replaced by K$^+$. This is originated from their different Born solvation energy. K$^+$ has a lower Born solvation energy in the micelle due to its larger ion radius in comparison with Na$^+$. Thus, K$^+$ is more preferable to be distributed in the vicinity of the micelle, which leads to a strong screening effect on the negative charge of SDS. Effectively, this makes SDS less hydrophilic and reduces its CMC. The reduction of CMC is more pronounced in MgCl$_2$ solution, because the negative charge of SDS is highly screened by the divalent Mg$^{2+}$. We need to emphasize that we do not perform extra data fitting in the calculation of different salt solutions. All the F-H parameters are the same as those calibrated in the NaCl solution. The Born radii of different ions are adopted from literature. Without any extra fitting parameters, our theory can fully capture the specific ion effect on CMC in simple salt solutions.

\subsection{3.3 
Sodium Lauryl Ether Sulfate (SLES) Surfactants}
\begin{figure}[H]
\centering
\includegraphics[width=0.4\textwidth]{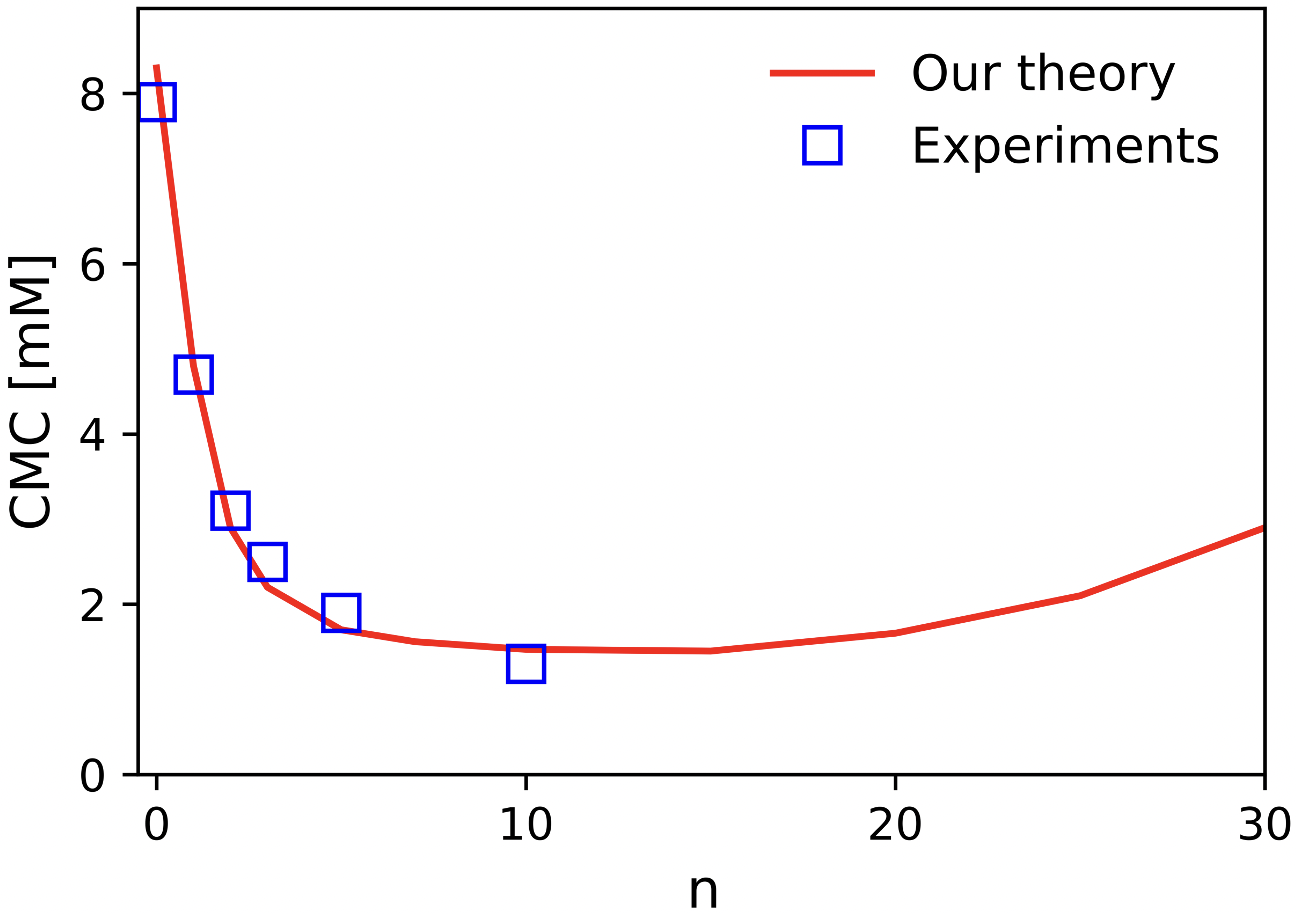}
\caption{
Comparison of the CMC predicted by our theory and experimental data reported in Ref. \cite{Del_Regno_2021} for SLES surfactants with various number of oxyethylene groups ($n$) in salt-free solutions. The Flory-Huggins parameters $\chi_{\rm CW}=2.47$, $\chi_{\rm EW}=0.09$ and $\chi_{\rm CE}=0.25$ are taken from the C$_m$E$_n$ system as in Fig. \ref{CMC_CmEn}. The values of $\chi_{\rm CS}=3.5$ and $\chi_{\rm SW}=-2.5$ are taken from the SDS system in Fig. \ref{CMC_SDS}. $\chi_{\rm ES}=-0.8$ in the SLES system is obtained from the best fit to the experimental data. 
}
\label{CMC_AES}
\end{figure}

Sodium lauryl ether sulfate (SLES) surfactants have $n$ additional oxyethylene groups between the hydrophobic dodecyl chain and the charged sulfate group, which can be considered as the combination of the C$_m$E$_n$ surfactants and the SDS surfactant. If $n=0$, SLES recovers SDS. As a milder alternative to SDS, SLES surfactants are more versatile, eco-friendlier, with better foam stability and hard water tolerance. \cite{BARRACARACCIOLO201794,Sasi2021}
Besides, SLES is flexible in formulation: the number of oxyethylene groups $n$ can be used to adjust its
micellization and interfacial properties. Figure \ref{CMC_AES} plots the CMC of SLES as a function of $n$ in salt-free solutions, which shows a quantitative agreement with the reported experimental data. In our calcuclation of the SLES surfactnts, the values of the F-H parameters $\chi_{\rm CW}$, $\chi_{\rm EW}$ and $\chi_{\rm CE}$ are taken from the C$_m$E$_n$ system, whereas the values of $\chi_{\rm SW}$ and $\chi_{\rm CS}$ are taken from the SDS system. The only unknown parameter $\chi_{\rm ES}$ in the SLES system is obtained by the best fit to the experimental data which leads to $\chi_{\rm ES}=-0.8$.   

Figure \ref{CMC_AES} shows a non-monotonic dependence of the CMC on the number of oxyethylene groups $n$. This nonmonotonic behavior is a result of the competition between the charged sulfate group and the charge-neutral hydrophilic oxyethylene groups. For small $n$, the CMC decreases rapidly as $n$ increases. The hydrophilicity in this regime is dominated by the charged sulfate group, which is expelled from the micelle core as $n$ increases. This reduces the effective hydrophilicity and hence leads to a reduction of CMC. The CMC reaches a minimum at $n=20$. Subsequently, the CMC turns to a slow increase as we add more oxyethylene groups because of their dominant contribution to the hydrophilicity in this regime ($n>20$).

\begin{figure}[H]
\centering
\includegraphics[width=0.4\textwidth]{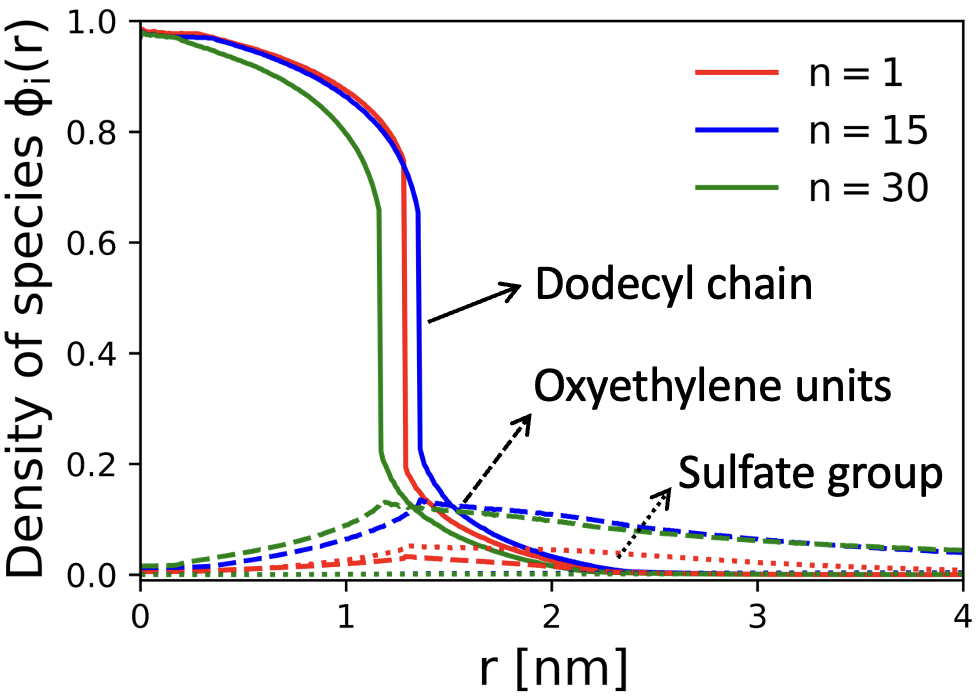}
\caption{
Microstructures of micelles formed by SLES surfactants with various number of oxyethylene units ($n$) in 0.2mM NaCl solution. Solid, dashed, and dotted lines represent the dodecyl chain, oxyethylene units, and sulfate group, respectively.
}
\label{Profile_AES}
\end{figure}

Similar non-monotonic $n$ dependence can also been found in the micelle structures. Figure \ref{Profile_AES} plots the density distributions of the representative micelles at CMC with the critical aggregation number. The micelle core becomes larger as $n$ increases from 1 to 15, caused by the expelling of the sulfate group from the micelle core. This reduces the repulsive charge interactions, leading to a decrease of $a_{H}$ and hence a increase of $P$. In contrast, the micelle core shrinks as $n$ further increases to 30 due to the dominant role of oxyethylene groups over the sulfate group in the regime of large $n$. Our results indicate the $n$ is a critical structure parameter for the design of SLES, where a continuous shifting from the ionic to nonionic surfactants can be realized.

\section{4. Conclusions}

We develop a self-consistent field theory to quantify the CMC of both nonionic and ionic surfactants. Our theory has the advantages over the existing work in the following aspects. (1) The calculation of the structure and free energy of a single micelle is systematically incorporated into the dilute solution thermodynamics, which allows us to unify the study of CMC, micellar structure and kinetic pathway of micellization in a unified framework. (2) The long-range electrostatic interactions are decoupled from the short-range van der Waals interactions and explicitly treated. This enables us to capture a variety of salt effects
like counterion binding, salt concentration dependence
and specific ion effect. (3) Intrinsic molecular features of the surfactant such as composition, architecture and charge
pattern are also explicitly accounted for, which makes our theory applicable to a broad range of surfactants and facilitates the direct comparison with experiments.

To validate the effectiveness and versatility of our theory, we apply it to three types of commonly
used surfactants, polyoxyethylene alkyl ethers
(C$_m$E$_n$), sodium dodecyl sulfate (SDS) and sodium lauryl ether sulfate (SLES). We calculate the CMC of C$_m$E$_n$ for a wide span of composition parameters, $m=8 \sim 16$ and $n=2 \sim 15$. With all the Flory-Huggins interaction parameters taken from the literature, our theoretical predictions are in quantitative agreement with the reported experimental data for a broad range of CMC from $10^{-6}$ to $10^{-2}$M. For the ionic SDS surfactant, we quantitatively capture the decrease of CMC as salt concentration increases. The microstructure of the micelles is also depicted in terms of the species distribution, which enables us to evaluate the counterion binding fraction. Our theory captures both the specific cation effect and the specific anion effect on the CMC. The molecular origin of the salt effect is elucidated as an interplay between the ion solvation and electrostatic screening. Furthermore, for SLES surfactants, we find a non-monotonic dependence of both the CMC and micelle size on the number of oxyethylene groups. Using only the minimum number of fitting parameters, our theoretical predictions are in quantitative agreement with the experimental data reported in the literature. 

The theory developed in this work provides an effective computational platform to study the micellization of surfactants with a variety of intrinsic molecular structures and under different external conditions. A continuous expansion of the library from calibrated surfactants to new ones can be achieved based on our parameterization strategy. Although we focus on the micelles formed by a single type of surfactant in the current work, our theory can be easily generalized to model surfactant mixtures where both the aggregation number and composition of the micelles need to be tracked in the SCFT calculation. The future work will also incorporate the same molecular model into a calculation at the water-oil interface. This will enable us to study the micellization and interfacial tension of surfactants in a unified framework.

\section{Acknowledgements}

This work was supported by Procter \& Gambel Company through Research Agreement 2023-053724. The research used the computational resources provided by the Kenneth S. Pitzer Center for Theoretical Chemistry at University of California Berkeley.



\end{document}